\documentclass[fleqn,10pt]{wlscirep}

\newcommand{\vp}{\varphi}
\newcommand{\w}{\omega}
\newcommand{\W}{\Omega}
\newcommand{\e}{\varepsilon}
\newcommand{\ii}{\mathrm{i}}
\newcommand{\dd}{\mathrm{d}}

\title{Real-time estimation of phase and amplitude with application to neural data}

\author[1,*]{Michael Rosenblum}
\author[1]{Arkady Pikovsky}
\author[2]{Andrea A. K\"uhn}
\author[2]{Johannes L. Busch}
\affil[1]{Department of Physics and Astronomy, University of Potsdam, 
Karl-Liebknecht-Str. 24/25, D-14476 Potsdam-Golm, Germany}
\affil[2]{Movement Disorders and Neuromodulation Unit, Department of Neurology, 
Charit\'e – Universit\"atsmedizin Berlin, Charit\'eplatz 1, D-10117 Berlin, Germany}
\affil[*]{corresponding author: mros@uni-potsdam.de}

\begin{abstract}
Computation of the instantaneous phase and amplitude via the Hilbert Transform is a powerful tool of data analysis. This approach finds many applications in various science and engineering branches but is not proper for causal estimation because it requires knowledge of the signal's past and future. However, several problems require real-time estimation of phase and amplitude; an illustrative example is phase-locked or amplitude-dependent stimulation in neuroscience.  In this paper, we discuss and compare three causal algorithms that do not rely on the Hilbert Transform but exploit well-known physical phenomena, the synchronization and the resonance. After testing the algorithms on a synthetic data set, we illustrate their performance computing phase and amplitude for the accelerometer tremor measurements and a Parkinsonian patient's beta-band brain activity.
\end{abstract}

\begin{document}

\flushbottom
\maketitle
\thispagestyle{empty}

\section*{Introduction}
Estimation of the instantaneous phase and amplitude is essential for electrical and mechanical engineering, synchronization studies of oscillatory systems of different nature,
time series analysis of physiological data, and, in particular, for  
neuroscience~\cite{Feldman-11,Liu-12,Drongelen-2006,Pikovsky-Rosenblum-Kurths-01}. 
A special application is developing efficient sensing algorithms for adaptive deep brain stimulation, a recent advancement of a widely used treatment option for Parkinson's disease and other neurological disorders ~\cite{Kuehn-Volkmann-17}.
One of the directions in this development is to adjust stimulation parameters according 
to a peripheral or neurophysiological signal's phase and amplitude computed on the 
fly~\cite{Rosin-11,Little-13,Cagnan_at_al-13,Rosa_et_al-15,Cagnan_at_al-17,%
Holt1119,McNamara2020.05.21.102335}. 
The popular approach to phase and amplitude estimation is to exploit the analytic signal approach based 
on the Hilbert Transform (HT) or, equivalently, the wavelet transform with a complex wavelet~\cite{Gabor46,Rabiner-Gold-75,Boashash-92,Feldman-11}.
However, this widely-used tool is non-causal and therefore not appropriate 
for real-time analysis, whereas causal estimation of phase and amplitude is 
often crucial for closed-loop control of complex systems.
Despite several attempts to adapt the HT for a causal 
measurement~\cite{Chen_et_al-13,Schreglmann-21}, 
the reliable and fast estimation of phase and amplitude of real-world signals remains challenging.

In this study, we do not rely on the Hilbert Transform. Instead, we follow another development line and extend our previous 
approach~\cite{Tukhlina-Rosenblum-Pikovsky-Kurths-07,Montaseri_et_al-13}
for real-time estimation based on the oscillation theory and nonlinear dynamics. 
The main idea is as follows. Suppose we have an oscillator, e.g., an electronic circuit, which amplitude and phase we can monitor. Next, suppose we feed our measurement to this oscillator. We choose the oscillator so that there is a one-to-one correspondence between the oscillator's phase and amplitude (what we can monitor) and the measured signal (what we want to determine). If this correspondence is achieved, we recompute the monitored state of the oscillator into desired quantities. Thus, the oscillator acts as a measuring device. Indeed, we will not exploit a physical oscillator but use a simple computer program that simulates the oscillator's dynamics.  
In our approach, we rely on two well-known physical effects: linear resonance and synchronization. We present and compare three techniques that provide a fast 
estimation of phases and amplitudes,
using only the past of the time series. We test the algorithms on model data and apply them to neural time series. Namely, demonstrating our approach's efficiency, we causally estimate the phase and amplitude of the accelerometer tremor measurements and Parkinsonian patients' beta-band brain activity and compare the results with non-causal HT-based analysis.

\section*{Results}
\subsection*{Hilbert Transform vs causal estimation}
A common practice in extracting the amplitude and the phase of an 
 input signal $s(t)$ is based on application of the Hilbert transform. 
 The HT analysis provides proper estimates for the instantaneous phase of $s(t)$ 
 for narrow-band one-component signals with slowly varying amplitude and frequency.  
 We emphasize that, though formally one can compute the HT for an arbitrary signal $s(t)$, 
 not in all cases the extraction of the amplitude and the phase will lead to reasonable results.
We also stress that HT is a non-causal operation: to extract features at time instant $t'$, one has to know both the signals' past $t<t'$ and future $t>t'$. For a further detailed discussion of the HT's properties and practical implementation, see Methods. 

The main goal of this paper is not to extend HT approach to complex signals 
(see \cite{Kralemann_et_al-13,Gengel-Pikovsky-19} for examples of such an extension), 
but to provide and explore causal alternatives to the HT approach.
In what follows, we use the HT-based amplitude $a_H$ and the HT-based phase $\vp_H$ as 
a ``gold standard'' for testing our algorithms. Strictly speaking,  this is reasonable for narrow-band signals only. 
Definition and determination of the proper phase and amplitude for a complex signal represent a challenging 
problem that we do not address here. Instead, we use an operational approach: a proper
amplitude should correctly
represent an envelope of the signal; a proper phase should gain $2\pi$ at each oscillatory cycle.
Below, we present three techniques for causal computations and compare the computed 
amplitudes and phases with non-causally obtained $a_H,\vp_H$.

\subsection*{Causal estimation techniques}
\subsubsection*{Phase locking approach}
The first technique exploits the ideas from the synchronization theory. It is well-known that a force $s(t)$ 
acting on a limit-cycle oscillator can entrain (lock) it. It means that the oscillator's frequency becomes equal to 
that of the force, and their phases differ by a constant. Thus, the phase of the locked limit-cycle oscillator
will correspond to the phase of the signal. 
For our purposes, it is helpful to use the simplest oscillator model, the so-called phase oscillator.  To ensure the phase-locking to the force, we have to adjust the oscillator's frequency to the signal's frequency. We assume that we do not know the latter 
\textit{a priori}, but can only roughly estimate it. 
We propose a simple approach that starts with this estimate and automatically tunes the ``device's'' frequency to 
ensure the locking and thus provides the instantaneous phase $\vp_L(t)$.  The amplitude is not determined with this approach.
One can treat the suggested scheme as a 
software implementation of a phase-locked 
loop~\cite{Best-84}. Technically, the algorithm reduces to solving differential equation 
incorporating measured data given at discrete time points; for details of the 
technique and its implementation, see Methods. 

\subsubsection*{Nonresonant linear filter}
The second technique relies on the resonance effect.  Our measuring ``device'' consists now of two linear 
damped oscillators.  The oscillators' frequency is much larger than the frequency of the signal, i.e., the system is 
far from resonance. We choose the damping parameters to ensure that (i) phase of the first linear oscillator 
equals that of the input and that (ii) amplitude of the second one and the input relate by a known constant 
multiplicator. The technique yields both phase $\vp_N(t)$ and amplitude $a_N(t)$, where the index $N$ stands 
for ``non-resonant''. 

\subsubsection*{Resonant linear filter}
Our third approach adopts the technique used for model studies in our previous publications~\cite{Tukhlina-Rosenblum-Pikovsky-Kurths-07,Montaseri_et_al-13}. 
The corresponding ``device'' consists of a linear oscillator in resonance with the measured signal and of an 
integrating unit. It also provides both the phase and the amplitude that we denote as $\vp_R$  and $a_R$, 
respectively. 
The method exploits the known relation between the resonant oscillator's phase and amplitude 
and those of the input. Additionally, the resonant oscillator acts as a bandpass filter for experimental data. 

Technically, both latest techniques require the numerical solution of linear 
differential equations with the input signal $s(t)$. 
We present a detailed description of the techniques and the developed numerical schemes 
in the Methods section. Like the phase-locking algorithm, both techniques include an automated frequency-tuning algorithm to adjust the systems 
to the a priori unknown signal's frequency. 

\subsection*{Testing the algorithms}

\subsubsection*{Artificial data}
\begin{figure}[thb!]
\centerline{\includegraphics[width=0.6\textwidth]{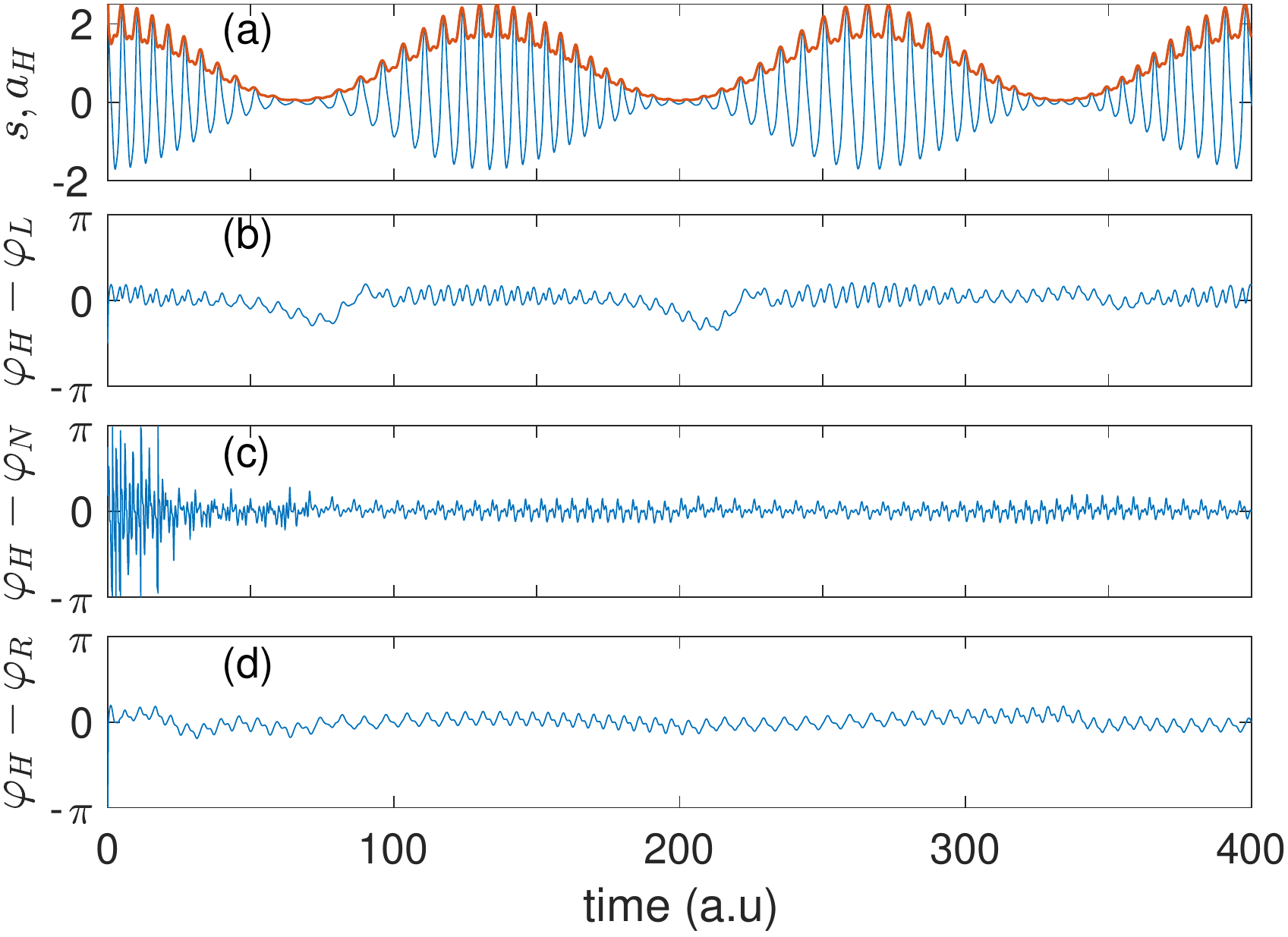}}
\caption{(a) The test signal $s$ according to Eq.~(\ref{eq:testsig}) and its 
Hilbert amplitude $a_H$ (red); one can see that $a_H$ does not represent 
a good envelope for $s$. 
Panels (b, c, d) show the difference between the Hilbert phase $\vp_H$ and 
causally estimated phases ($\vp_L$, $\vp_N$, and $\vp_R$ are obtained by 
means of the locking-based technique, non-resonant and resonant oscillator, 
respectively). Notice that in (c) we show $(\vp_H-\vp_N)\mod 2\pi$: within the 
first 20 time units the phase difference decreases to $-14\pi$ until it saturates
and oscillates around $(\vp_H-\vp_N)\mod 2\pi=0$.
}
\label{fig:test1}
\end{figure}

\begin{figure}[thb!]
\centerline{\includegraphics[width=0.6\textwidth]{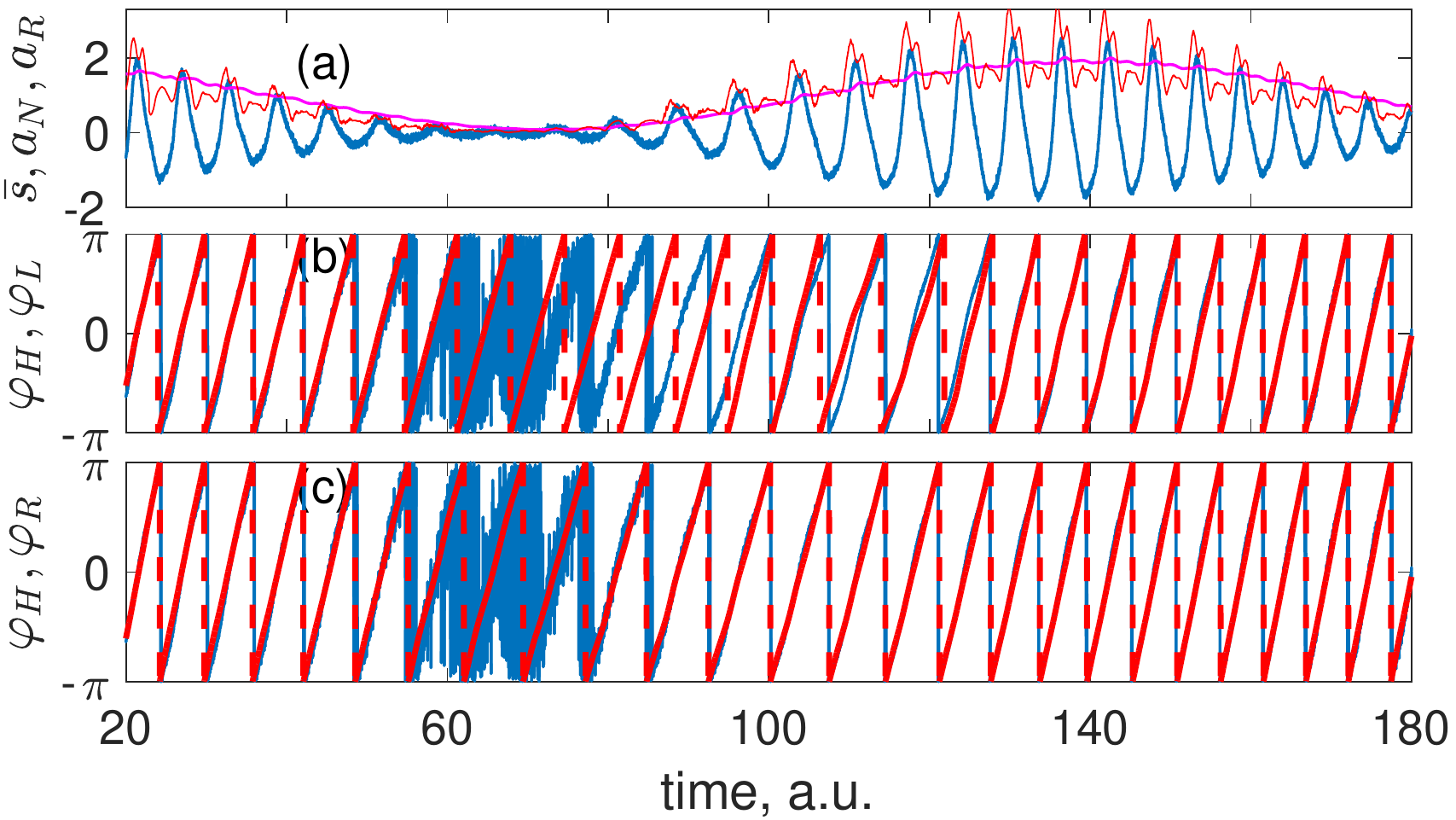}}
\caption{(a) Short epoch of noisy data $\bar s$ (blue) and causally obtained amplitudes
$a_N$ (red)  and $a_R$ (magenta); the latter provides the most smooth envelope 
(cf. $a_H$ in Fig.~\ref{fig:test1}a).
When the signal's amplitude is very small, the noise dominates and phase 
determination becomes complicated. The HT approach fails here (blue line in (b,c)), 
while both locked and resonant oscillator ``devices'' provide reasonable results.
The performance of the non-resonant technique is poor: when the amplitude nearly vanishes, the phase estimated by this technique (not shown) is not better than the Hilbert phase.
}
\label{fig:test2}
\end{figure}

First, we test our approach on artificial data. All algorithms work well with simple narrow-band signals like slowly modulated sine waves. So, we do not show these 
results and proceed with a more complicated case. The signal is 
\begin{equation}
s(t)=\left [1+0.95\cos(\W_1 t)\right]\cdot\left[ \cos(\psi(t))+
0.2\cos(2\psi(t)+\pi/6)+0.1\cos(3\psi(t)+\pi/3)\right]  \;,
\label{eq:testsig}
\end{equation}
where
\[
\psi(t)=t+5\sin(\W_2 t)\;,
\]
its waveform has three harmonics. The signal is amplitude- and phase-modulated, 
see Fig.~\ref{fig:test1}a; it is sampled with $\Delta=0.01$ to yield 
$s(\Delta k)=s_k$. The frequencies are $\W_1=\sqrt{2}/30$, $\W_2=\sqrt{5}/60$.
Another test signal is $\bar s_k=s_k+\xi_k$, where $\xi_k$ are Gaussian random 
numbers with zero mean and standard deviation $0.05$, 
see Fig.~\ref{fig:test2}a.
For testing all three algorithms, we set the initial frequency of the device to be $10\%$ higher than the actual value; in this way, we imitate the imprecision on initial frequency estimation. 
The other parameters are given in the Methods. 

\paragraph*{Phase-locked oscillator.}
Figure~\ref{fig:test1}b shows that after a short transient, the phase determined by the ``device'' becomes close to the Hilbert phase. 
The difference is due to the presence of the harmonics 
(for a mono-component amplitude-modulated signal, 
the phase difference is very small, its standard deviation is $0.03$). 
The locking-based measurement's advantage becomes evident in the noisy data, as expected for a phase-locked loop approach: here, the real-time estimation provides a reasonable phase even at low amplitudes, where the noise dominates, see Fig.~\ref{fig:test2}a,b. 

\paragraph*{Non-resonant linear oscillator (filter).}
This approach works well in noise-free case (Fig.~\ref{fig:test1}c). 
Initially, the difference with $\vp_H$ 
grows rapidly due to decaying high-frequency oscillation with the natural frequency 
$\w$ of the system. 
(We remind that frequency $\w$ of the non-resonant oscillator is much larger than 
the input's frequency.) 
Nevertheless, after the transient, the performance is comparable or even better 
than that of two other techniques.
In the presence of noise the performance is poor because noisy perturbations excite high-frequency oscillation with the frequency $\w$. 
The amplitude estimation is slightly worse than that via HT, see Fig.~\ref{fig:test2}a. 
Thus, for processing without a filter, this technique is not optimal.

\paragraph*{Resonant linear oscillator (filter).}
This technique also demonstrates efficient phase estimation, 
see Fig.~\ref{fig:test1}d for the noise-free data and Fig.~\ref{fig:test2}c for the 
noisy case. 
Actually, the results for the latter are hardly distinguishable from those 
in Fig.~\ref{fig:test2}b, 
though the frequency adaptation of the resonant oscillator is faster.  
The technique also provides the instantaneous amplitude, see Fig.~\ref{fig:test2}a. 
Notice that the Hilbert amplitude $a_H$ 
for the multi-component signal $\bar s$ is not the perfect envelope.
The real-time amplitude $a_R$ is not perfect either, but is much more smooth 
than $a_H$.

\subsubsection*{Estimation without an additional filter: tremor data}
\begin{figure}[thb!]
\centerline{\includegraphics[width=0.53\textwidth]{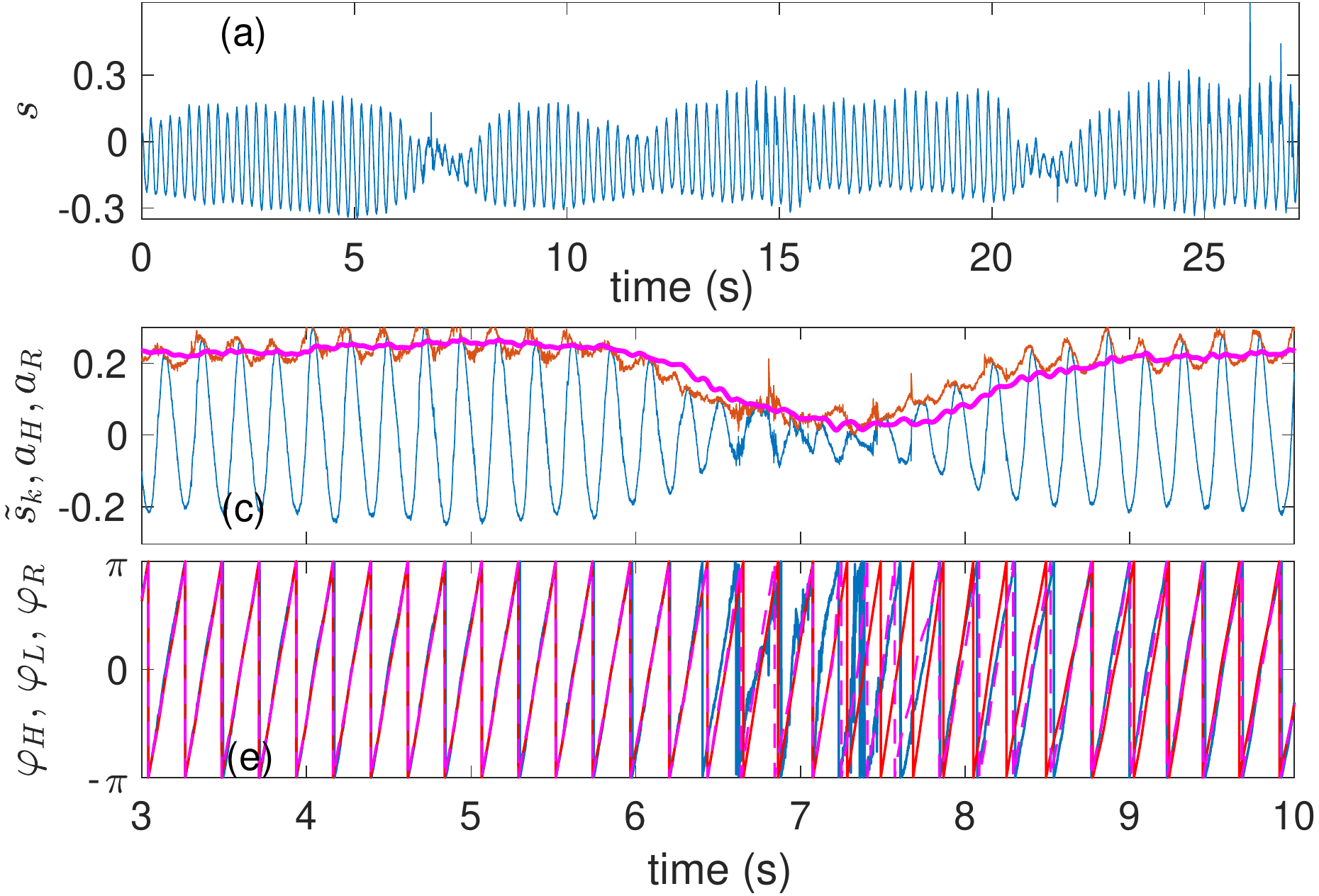}
\includegraphics[width=0.25\textwidth]{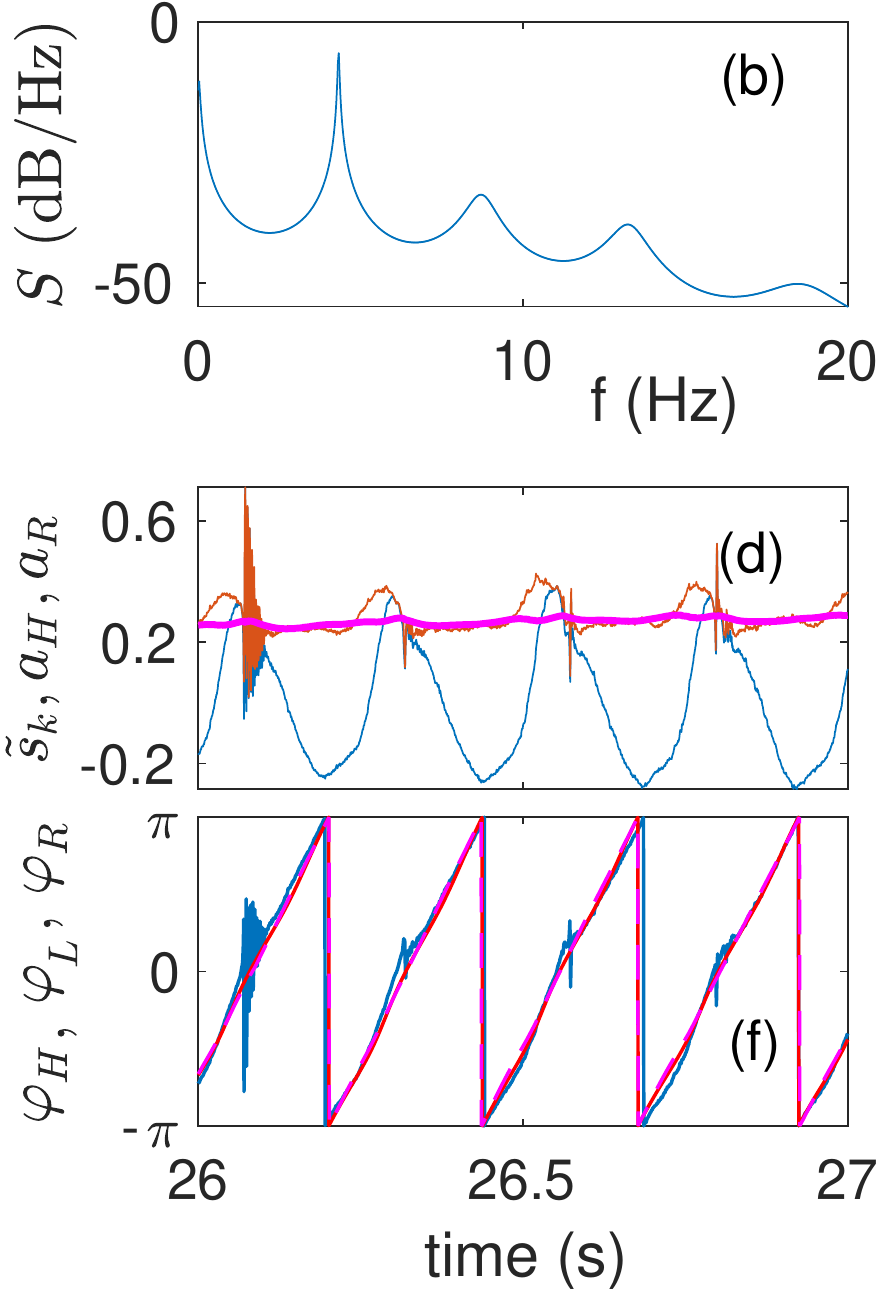}}
\caption{The raw tremor data (a) and its power spectral density $S(f)$, computed by the Burg algorithm (b). 
Panels (c,d) show some epochs of $\tilde s_k$ (the original signal with the constant component and baseline fluctuation removed on the fly), the Hilbert amplitude $a_H$ (red)  and the causal amplitude $a_R$ (magenta). 
The latter provides a smooth envelope and does not reflect the outliers (d).
Panels (e,f)  demonstrate the Hilbert phase $\vp_H$ (blue) and causal phases  $\vp_L$ (red)  
and $\vp_R$ (magenta). For epochs with vanishing amplitude,
causal phases do not exhibit noisy jumps as the Hilbert phase.  
The causal phases are stable with respect to outliers (f). 
}
\label{fig:2}
\end{figure}

Tremor data was acquired from a patient with essential tremor using a 3D accelerometer (TMSi, Oldenzaal, The Netherlands) attached to the right index finger. The signal was sampled at 2048 Hz sampling rate using a Porti amplifier (TMSi). The patient stretched their arms out in front of their chest holding a bottle in order to provoke postural tremor. Only data from one axis was analyzed.
The tremor time series (Fig.~\ref{fig:2}a) 
is relatively simple: it resembles the test signal from the previous section. 
The tremor data is amplitude 
modulated, and its power spectrum also contains 
several harmonics  (Fig.~\ref{fig:2}b).
However, the real-world signals are naturally more complicated than the simple 
test data series Eq.~(\ref{eq:testsig}). 
Analyzing them, one frequently faces additional difficulties. 
For the tremor signal under consideration these features are:  a drift of the baseline; presence of 
epochs when the amplitude 
vanishes; and outliers.  
To cope with the baseline fluctuation, we exploit the following causal detrending algorithm. For each new measurement point, we remove the mean value computed over several previous cycles.  For the sake of computational speed, we update this 
value several times per period. We  denote the detrended signal by $\tilde s_k$ (Fig.~\ref{fig:2}c).

Figure~\ref{fig:2}c compares the Hilbert amplitude $a_H$ with the causally computed $a_R$. 
Like in the case of test data, the causally computed amplitude $a_R$ is a better estimation of the envelope 
than $a_H$  is. Moreover, computation of $a_R$ is stable with respect to outliers (Fig.~\ref{fig:2}d).
Next, in Fig.~\ref{fig:2}e,f we show the Hilbert phase $\vp_H$ along with two causally-obtained 
phases $\vp_L$ and $\vp_R$.
We see that for large-amplitude oscillation, all the phases practically coincide. 
When the amplitude almost vanishes, the ``device'' of the locking-based method falls out of synchrony and makes one cycle less.
On the contrary, the resonant-oscillator phase $\vp_R$ and the Hilbert phase reveal the same
number of cycles. 
We emphasize, that both causal techniques are not sensitive to the outliers in the original data,
while the Hilbert approach is  (see  Fig.~\ref{fig:2}f). Finally, we mention that the non-resonant 
oscillator technique works poorly for the unfiltered tremor data.  Only if we use the bandpass filter 
$4.5\pm 2$~Hz, then this technique works perfectly.

\subsubsection*{Wide-band signal: beta-band brain activity}
Elevated beta-band activity has been established as a marker for rigidity and bradykinesia in patients with Parkinson's Disease \cite{kuhnReductionSubthalamic352006}. As such, it has been employed in several studies of adaptive deep brain stimulation aiming for automatic adjustment of stimulation parameters in response to beta-band amplitude \cite{Little-13, Rosa_et_al-15}. Here we recorded LFP data from a patient with Parkinson's Disease on medication, 2 days after surgery for deep brain stimulation with electrode cables being externalized. LFPs were acquired from the left subthalamic nucleus via a 4-contact electrode (Model 3389, Medtronic,  Minneapolis,  USA) using a D360 amplifier (Digitimer Ltd., Welwyn Garden City, UK) and were digitized at 1 kHz sampling rate with a 1401 analog-to-digital converter (CED Ltd., Cambridge, UK). A bipolar referencing scheme was adopted, with the signal shown here originating from contacts one to two. During the recording session, the patient was comfortably seated and was asked to rest quietly.

\begin{figure}[thb!]
\centerline{\includegraphics[width=0.6\textwidth]{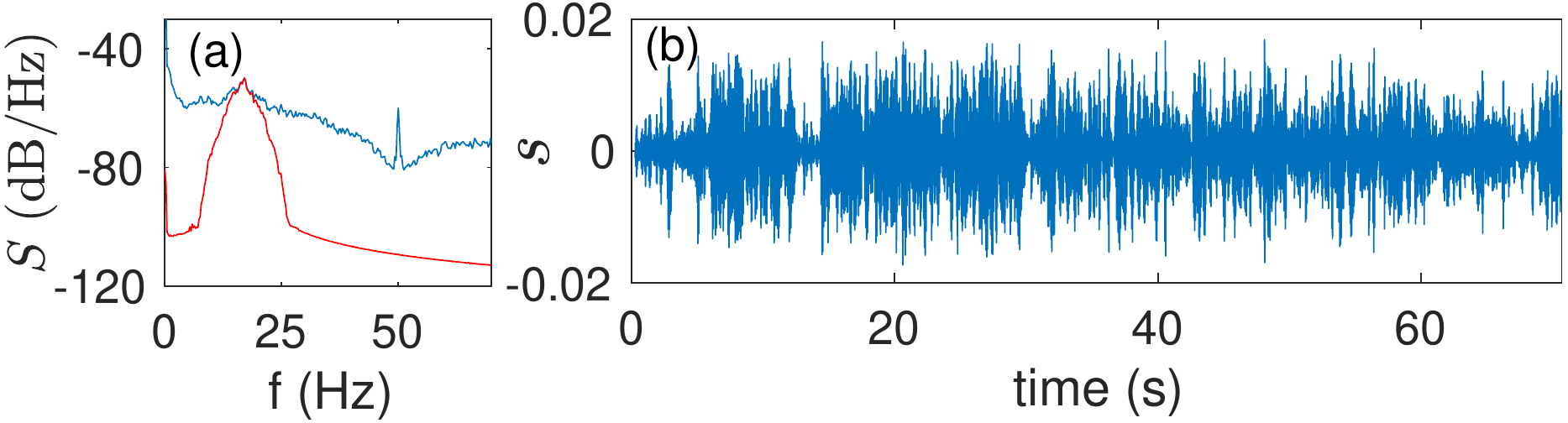}}
\centerline{\includegraphics[width=0.6\textwidth]{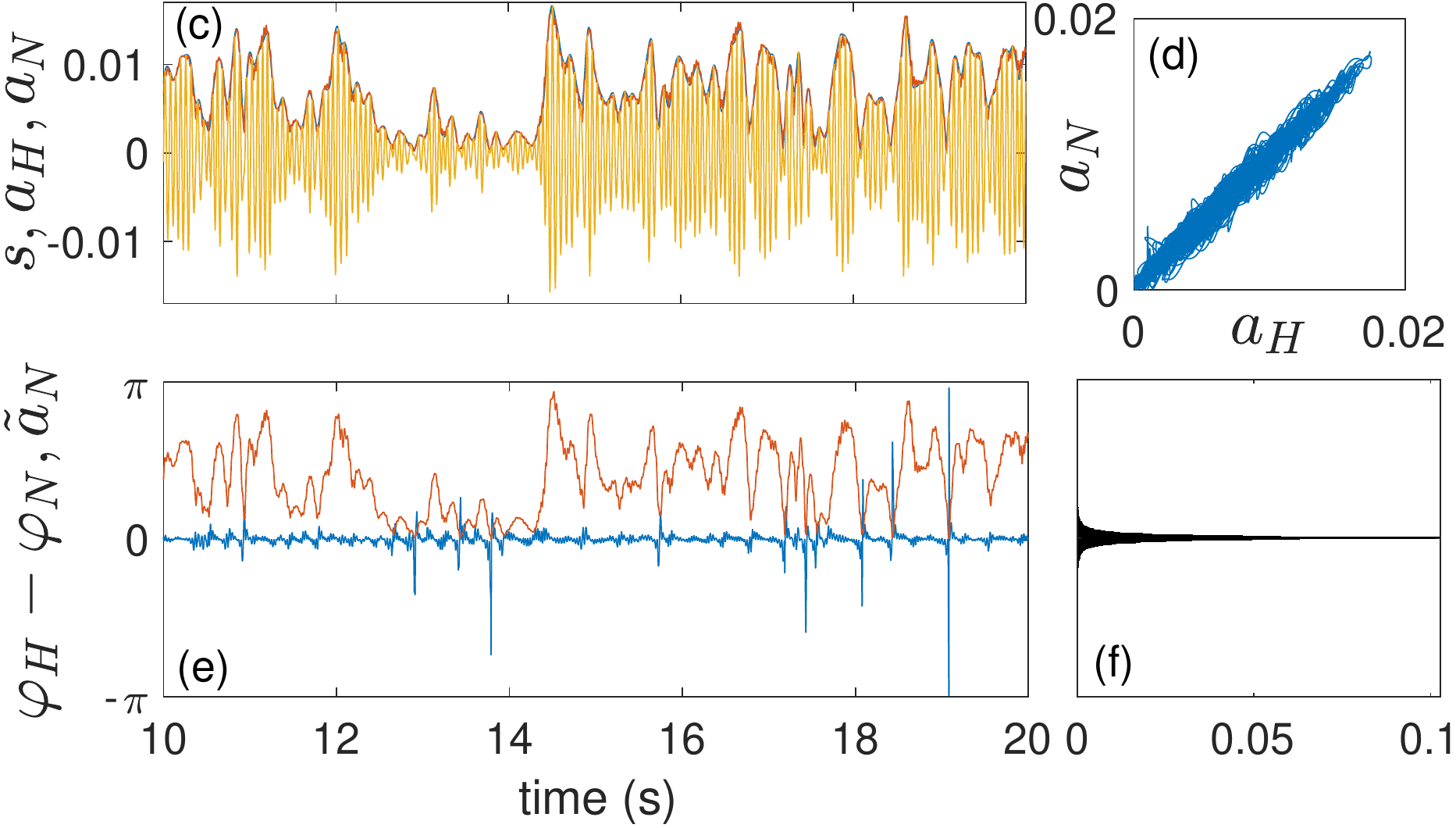}}
\caption{Beta-band brain activity data. (a) Power spectral density $S(f)$ of the raw (blue) and filtered (red) series. (b) Bandpass filtered series (arbitrary units).
(c) A short epoch of the filtered series (yellow) and two their envelopes computed via the 
Hilbert Transform (blue) and by means of the non-resonant oscillator (red). The envelopes
practically overlap, as is also illustrated by panel (d). Panel (e) demonstrates  
the difference between the Hilbert phase $\vp_H$ and the real-time phase $\vp_N$
along with the amplitude: one can see 
that the phase difference is negligible  when the amplitude does not vanish and is 
not small only when amplitude is on the level of noise so that the phase is not 
well-defined; for these amplitude values, reliable phase determination is not possible.
(Here $\tilde a_N$ is the amplitude rescaled for better visibility.) 
The probability distribution of the phase difference in panel (f) confirms practical coincidence of $\vp_H$ and $\vp_N$. 
}
\label{fig:6}
\end{figure}

The analysis of the beta-band brain activity requires a bandpass filter preprocessing.
We use a simple FIR filter with bandwidth $17\pm 4$~Hz ($281$-point filter was generated by the Matlab {\tt fir1} function).  
A causal filtration introduces a delay, but this is a necessary price to be paid. 
Our tests with this signal demonstrate that the non-resonant-oscillator technique outperforms the other two that fail because of the substantial variation of the signal's amplitude.
Noteworthy, due to the bandpass, the approach works without the baseline correction and the 
frequency adaptation. Hence, the algorithm implementation requires only a few lines of code. The results 
are illustrated by Fig.~\ref{fig:6}.  

\section*{Discussion}
To summarize, in this paper, we considered the problem of causal instantaneous amplitude and frequency estimation.
Contrary to attempts to adjust the inherently non-causal analytic signal approach, we used oscillation theory ideas. 
 We presented and tested three algorithms. The first one exploits the phase-locking property of nonlinear limit-cycle oscillators, while the other two rely on linear resonance. Technically, we encountered solving differential equations incorporating experimental data; we suggested efficient numerical schemes to solve them. Below, we discuss the advantages and disadvantages of all three techniques. Since these techniques aim at real-time applications, we pay special attention to their computational efficiency.

The locking-based technique is similar to the phase-locked loop approach and provides the phase only. It works well for relatively narrow-band signals like the tremor data and does not require an additional bandpass filter. It can easily incorporate the frequency adaptation that makes the algorithm able to cope with the signal's frequency's slow variation. The technique does not work well with the signals like beta-band activity, where both the frequency and the amplitude vary relatively fast; this results in frequent 
loss of synchrony. The algorithm is computationally efficient: the only demanding operations for a new phase value are several computations of the sine function.  

As an advantage of the resonant oscillator technique, we mention that it provides both the phase and the amplitude. It is stable to high-frequency noise because the resonant oscillator acts as a weak bandpass filter. However, it requires baseline correction because the integrator unit enhances the low-frequency perturbations. The implementation described in the Methods section is very efficient, provided the oscillator's frequency does not change. Adaptation of the frequency to that of the signal requires recomputation of algorithm's coefficients what slightly reduces the efficiency.

If extraction of the rhythm of our interest requires a bandpass filter, the non-resonant oscillator approach is the best choice. Using the beta-band brain activity as an illustrative example, we demonstrated that this causal algorithm provides phase and amplitude very close to non-causally computed Hilbert Transform-based values. 
An essential advantage of the algorithm is its efficiency: computation of the new phase and amplitude values requires only several arithmetic operations and two function calculations. This property makes the algorithm especially appropriate for the real-time processing of high-frequency signals. 			

We emphasize that all three techniques are much faster than approaches based on the HT. 
For example, the technique introduced in \cite{Schreglmann-21} requires the computation of the direct and inverse Fourier transform in a running window plus additional filtration to compensate for the boundary effects and non-causal nature of the HT; altogether, this computation is about thirty times 
more demanding than an oscillator-based approach.  

Taken together, the techniques suggested here may provide useful means for real-time detection of phase and amplitude in the context of adaptive deep brain stimulation. For example, estimation of instantaneous amplitude has been implemented by interpolating beta-filtered and squared LFP data \cite{petrucciClosedloopDeepBrain2020}. However, this approach is not able to recover the signal's phase simultaneously. Additionally tracking phase may augment therapeutic benefit, though, by delivering stimulation pulses at specific points within the oscillatory cycle. This has been shown to effectively disrupt pathological oscillations \cite{McNamara2020.05.21.102335, Schreglmann-21, Cagnan_at_al-17, Holt1119}. So far, to our knowledge, adaptive stimulation algorithms integrating both the phase and amplitude of a signal have not been established yet.

Finally, we mention that our techniques do not cause any additional delay. The processing delay is only due to a causal filter if the latter is required, e.g., for brain activity data. To minimize the delay, one has to exploit specially designed filters  \cite{Smetanin_et_al-20}.

\section*{Methods}
In all cases 
the input is an oscillatory signal $s(t)$ is sampled with interval $\Delta$. Thus, available is a 
sequence $s(t_k)=s(k\Delta)=s_k$.

\paragraph{The non-causal Hilbert-Transform based approach.} In this approach one constructs the corresponding complex-valued analytic signal $Z(t)=s(t)+\ii \hat s(t)$, where 
$\hat s(t)=\pi^{-1}\text{P.V.}\int_{-\infty}^\infty\frac{s(\tau)}{t-\tau}\dd\tau$ is the HT of $s(t)$.
Obviously, the HT is a non-causal operation; HT of a finite-length time series yields spurious 
values for the boundaries. 
The absolute value and argument of $Z(t)$ provide the instantaneous Hilbert amplitude $a_H(t)$ 
and phase $\vp_H(t)$, respectively.  For a narrow-band one-component signals like $s(t)=A(t)\cos[\w(t)t]$, where $A(t)$, $\w(t)$ are slow functions of time, the analytic signal approach 
provides $a_H\approx A$ and $\vp_H\approx \int \w(t) \dd t$. 
For a discussion of the HT's practical implementation and technical hints we refer, e.g., to \cite{king_2009,Feldman-11,Pikovsky-Rosenblum-Kurths-01}. In a practical implementation either a discrete evaluation
of the integral is performed, or a discrete Fourier transform is used.

\subsection*{Causal estimation of phase and amplitude}
In these methods, we obtain the value of the instantaneous phase 
$\vp(t_k)=\vp_k$ and amplitude $a(t_k)=a_k$ by using only the current and the previous values of the signal, i.e., $s_k,\; s_{k-1},\; \ldots$. 

\subsubsection*{Measuring ``device'': phase-locked oscillator}
The synchronization theory says that an oscillatory force $s(t)$ acting on a limit-cycle oscillator 
can entrain it, if the frequency of the force is close to the natural
frequency of the limit-cycle oscillator. 
It means that the oscillator's frequency becomes equal to that of the force, 
and their phases fulfill the 
locking condition $\vp-\theta\approx\mbox{const}$, 
where $\theta$ and $\vp$ are the oscillator's and the signal's phases.
For our purposes, it is appropriate to use the so-called phase oscillator. 
Its forced dynamics is described by 
\begin{equation}
\dot\theta =\w-\e \sin\theta\cdot s(t)\;,
\label{eq:phosc}
\end{equation}
where $\e$ is a parameter that determines the coupling strength.
Consider first a harmonic force, $s=a\cos(\nu t)$.
If $\w=\nu$, then in the locked state $\vp\approx\theta$. 
However, $\nu$ is not known \textit{a priori}, but we assume that we can roughly 
estimate it. Let this initial guess be $\nu_0$. Thus, we set initially $\w=\nu_0$
and start our computation with this value. 
To adapt the phase oscillator to the signal, we estimate the 
frequency of the forced oscillator $\nu_e$ on the fly. (The index $e$ stands for ``estimated'').
To this end, for the time instant $t$, 
we take previously computed \textit{unwrapped} phases $\theta$ for the 
time interval $[t-T_e,t]$, where $T_e\sim 2\pi/\nu_0$ (approximately one or two cycles).
Assuming that within this time interval $\theta(t')=\theta(t-T_e)+\nu_e(t'-t+T_e)$, 
we compute the frequency $\nu_{e}$ via the linear fit.  
Next, we update the oscillator's autonomous frequency as
\begin{equation}
\w\quad \longrightarrow \quad \w+K(\nu_e-\w)\;,
\end{equation}
where $K$ is a constant update factor.
Adapting the frequency in this way, we ensure a transition from the 
unlocked to locked dynamics.
Performing frequency estimation several times per cycle, we successfully estimate 
the phase of signals with slowly drifting frequency.
We expect that this algorithm also works if the force's amplitude $a$ 
slightly varies with time (so that the oscillator still remains locked). 

One can treat the suggested scheme as a software implementation of a phase-locked 
loop~\cite{Best-84}, cf. also~\cite{Rosenblum-Pikovsky_et_al-02}.
Practically, we have to solve the differential Eq.~(\ref{eq:phosc}) numerically, 
whereas its right-hand side is known only in discrete time points $t_k$.
The easiest way is to exploit Euler's technique with the integration step $\Delta$ to
advance from the known value $\theta_k$ at $t_k$ to the new phase $\theta_{k+1}$
at $t_{k+1}=t_k+\Delta$, given a new measurement $s_{k+1}$. This technique may 
work properly if $\Delta$ is sufficiently small; otherwise, the numerical solution becomes unstable. 
(We recall that $\Delta$ is the fixed sampling interval that cannot be made arbitrarily small). Therefore, to advance the solution of Eq.~(\ref{eq:phosc}) from $\vp_k$ to $\vp_{k+1}$
given new measurement $s_{k+1}$, we use the parabolic approximation
of $s(t)$ 
on the interval $[t_k,t_{k+1}]$ and then exploit the standard Runge-Kutta algorithm.
(Notice that the Runge-Kutta step can be much smaller than $\Delta$.)
The coefficients of the parabolic fit are computed from $s_{k-1},s_{k},s_{k+1}$. 
Starting with some initial condition, e.g., $\theta_0=0$, we achieve,
after a short transient (see examples in Fig.~\ref{fig:test1}), the synchronous state 
where $\vp_k\approx\theta_k$.
Notice that the phase oscillator can be complimented by a low-pass filter:
\[
\dot\vp=\w+\e w\;, \qquad \tau\dot w +w=-s(t)\sin\theta\;, \qquad 
\]
then the scheme becomes the simplest traditional phase-locked 
loop~\cite{Best-84,Pikovsky-Rosenblum-Kurths-01}; for $\tau\to 0$ it reduces to Eq.~(\ref{eq:phosc}). 
This extension may be useful for the data with strong high-frequency noise 
but it does not show any advantages for the tremor data we use.

\subsubsection*{Measuring ``device'': non-resonant linear oscillator}
Our second measuring ``device'' is linear damped oscillator:
\begin{equation}
\ddot x+\alpha \dot x +\w^2 x= s(t)\;,
\label{eq1}
\end{equation}
where $\w$ and $\alpha$  are the frequency and the damping parameter of the oscillator, respectively. Suppose first that $s(t)$ is harmonic, $s(t)=a\cos(\nu t)=a\cos(\vp(t))$. 
The well-known stationary solution of linear Eq.~(\ref{eq1}) is 
$x=b\cos(\nu t+\beta)$, where $b=\frac{a}{\sqrt{(\omega^2-\nu^2)^2+(\alpha\nu)^2}}$ and
$\beta=\arctan\left[\frac{-\alpha\nu}{\omega^2-\nu^2}\right]$.
Thus, the forced system (\ref{eq1}) oscillates with the force's frequency $\nu$ and the amplitude $b$. The dependencies of the amplitude ratio $b/a$ and of the phase shift $\beta$ on the forcing frequency $\nu$ reflect the well-known resonance effect.  
Knowing the oscillator's state $x(t)=b\cos(\nu t+\beta)$, 
$\dot x(t)=-b\nu\sin(\nu t+\beta)$ 
we find the amplitude and phase of the external force:
\begin{equation}
a(t)=\sqrt{x(t)^2+[\dot x(t)/\nu]^2}\sqrt{(\omega^2-\nu^2)^2+(\alpha\nu)^2}\;,
\qquad
\varphi(t)=\arctan\left [\frac{-\dot x(t)}{\nu x(t)}\right ]-\beta\;.             
\label{eq2}
\end{equation}
Thus, if our ``device'' yields $x(t)$ and $\dot x(t)$, then we easily 
compute $a(t)$ and $\varphi(t)$, provided the frequency $\nu$ of the force 
is known. However, it is not known but can only be roughly estimated. Moreover,  generally, it varies with time. 
 Before we discuss how to cope with this fact, we mention how we practically obtain $x(t)$ and $\dot x(t)$. Here, we make use of the linearity of Eq.~(\ref{eq1}) and develop and exploit an efficient numerical scheme, presented below. 
With this scheme, starting with some initial conditions, e.g., $x_0=\dot x_0=0$, we obtain, after a short transient, the forced solution of Eq.~(\ref{eq1}) and compute 
$a(t_k)$ and $\varphi(t_k)$ from Eqs.~(\ref{eq2}).

\begin{figure}[thb!]
\centerline{\includegraphics[width=0.45\textwidth]{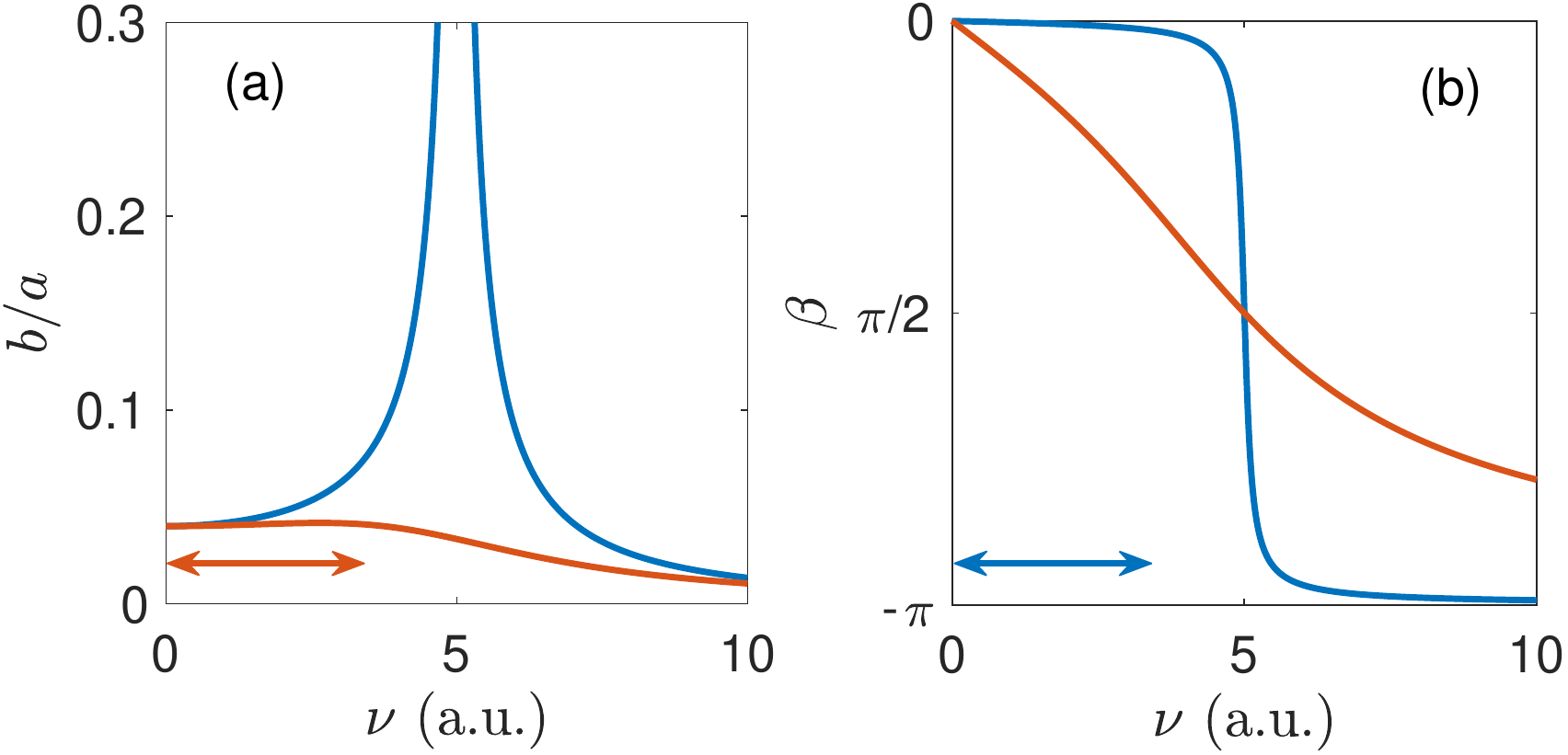}}
\caption{Resonance curves for amplitude (left) and phase shift (right)
for the linear oscillator with frequency $\w=5$.  
Blue and red curves correspond to the weakly ($\alpha=0.2$) and strongly 
($\alpha=6$) damped oscillator,  used for the phase and amplitude measurement, respectively
(these parameters were used to process the artificial data).
The domains where $b/a\approx \text{const}$ (for the red curve) and $\beta\approx 0$ (for the blue curve) are marked by red and blue arrows, respectively. 
These are intervals of signal frequency $\nu$ where the algorithm's performance is good. 
We see that $\w\approx 5\nu$ is a reasonable choice. 
}
\label{fig:1}
\end{figure}

Now, we discuss how to choose the parameters of the measuring oscillator. Recall that to compute $a,\vp$, we need the value of the signal frequency $\nu$. First, let us consider the amplitude measurement. Inspecting Fig.~\ref{fig:1}, we see that if we take a large value of $\alpha$ (strongly damped oscillator) and $\nu \ll \w$ then the ratio $b/a$ is practically independent on $\nu$. Thus, to compute the second square root in Eq.~(\ref{eq2}) we need only a very rough estimate of $\nu$. For the phase estimation, the damping parameter shall be different. Indeed, as follows from Fig.~\ref{fig:1}b, now we have to choose $\alpha$ to be small, then the phase shift $\beta\approx 0$ in a wide range of $\nu$ and the phase of the external force equals the phase of the oscillator. 
A reasonable choice is to take the oscillator's frequency $\w$ about five times larger 
than $\nu$. 
Thus, we neglect $\beta$ in the expression for $\vp$ and compute $\sqrt{(\omega^2-\nu^2)^2+(\alpha\nu)^2}$ in the expression for $a$ only once, using an initial guess $\nu_0$ for frequency. However, the terms $\sqrt{x(t)^2+[\dot x(t)/\nu]^2}$ and 
$\arctan\left (\frac{-\dot x(t)}{\nu x(t)}\right )$ essentially depend 
on $\nu$. Imprecise estimation of $\nu$ and/or its variation with time
results in spurious oscillations of the estimated amplitude and phase.
To remove these oscillations and to make the results less dependent on the initial frequency estimation, we improve this estimation by computing the signal frequency in real-time as discussed in the precious section. We start with an initial 
guess $\nu_0$ and begin the computation of $a(t)$, $\vp(t)$. 
After a transient time of the order of several cycles, we begin with the frequency 
estimation to obtain $\nu_e$. (The estimation is performed in exactly same way as 
for the phase-locked oscillator.)
We use $\nu_{e}$ instead of $\nu_0$ and, in this way, significantly improve the real-time computation of the amplitude and frequency. 
Performing frequency estimation several times per cycle, we track the slowly drifting signal frequency. 
(Notice that for the bandpass filtered signal used in the last example, 
the algorithms works well without any frequency correction. 
Here we set $nu$ equal to the center of the bandpass.) 

In summary, for the amplitude measurement, we need a strongly damped oscillator, while for the phase estimation, we need an oscillator with small damping. If we need both measurements simultaneously, then the solution is to exploit two ``devices'' concurrently. Though Eqs.~(\ref{eq2}) are derived for the harmonic force, we expect that they approximately hold for signals with a slow variation of the amplitude and frequency.  The presented numerical tests confirm this expectation.  

\subsubsection*{Measuring ``device'': resonant linear oscillator}
Here we adopt the technique used for model studies in our previous 
publications~\cite{Tukhlina-Rosenblum-Pikovsky-Kurths-07,Montaseri_et_al-13}. 
The device consists of a linear oscillator in resonance with the measured 
signal and an integrating unit:
\begin{align}
&\ddot{x}+\alpha\dot x +\w^2 x = s(t)\;,\label{eq:filter}\\
&\mu\dot{z} + z = \dot x \;.              \label{eq:integrator}
\end{align}
The role of the harmonic oscillator Eq.~(\ref{eq:filter}) is twofold. 
First, it acts as a bandpass filter and the 
damping factor $\alpha$ determines the width $\delta f\approx \alpha/2\pi$ 
of the bandpass.
Second, the harmonic oscillator yields signal $\dot x$ which phase is close to that of 
the input $x(t)$, provided the frequency $\w$ is close to the mean frequency of $s(t)$.
(This condition also ensures that the band-pass is centered at the frequency of the signal.)
Next, for $\mu\gg 1$ Eq.~(\ref{eq:integrator}) acts as the integrating unit. 
Its output is shifted by $\pi/2$ with respect to $\dot x$. 
It is useful to rescale $\dot x$, $z$ so that their amplitudes 
are close to that of $s(t)$. For this, we compute  
$u=\alpha\dot x$ and $w=\alpha\w_0\mu z$ 
and obtain the instantaneous phase and amplitude of $s(t)$ as
$\vp(t)=\arctan(w/u)$, $a(t)=\sqrt{u^2 + w^2}$. 

\subsection*{Solving the linear oscillator equation for discrete input signal}
Here we present the numerical scheme for solving Eq.~(\ref{eq1}),
adopted to the situation where the input signal is available at a finite sampling rate. 
Using the substitution 
$y=xe^{\gamma t}$, where $\gamma=\alpha/2$, one obtains
\[
\ddot y +\eta^2 y=s(t)e^{\gamma t}=S(t)\;, \quad\text{with  } \eta^2=\w^2-\gamma^2\;.
\]
Another standard substitution 
\begin{equation}
y=0.5(Ae^{\ii \eta t}+A^*e^{-\ii \eta t})\;, \qquad 
\dot y =0.5\ii \w (Ae^{\ii \eta t}-A^*e^{-\ii \eta t})\;,
\label{subst}
\end{equation}
where $A^*$ denotes complex conjugate of $A$,
yields $\dot A=-\ii \eta^{-1}S(t)e^{-\ii \eta t}$.
Integrating we obtain 
\[
A_{k+1}=A_k-\ii\eta^{-1}e^{-\ii \eta t_k}\int_0^\Delta S(t_k+\tau)e^{-\ii \eta\tau}\dd \tau
=A_k-\ii\eta^{-1} e^{-\ii \eta t_k}I\;.
\]
Next, locally interpolating the measured signal $s(t)$
by a parabola going through the points $s_{k-1},s_k,s_{k+1}$ we compute 
the integral $I=\int_0^\Delta S(t_k+\tau)e^{-\ii \eta\tau}\dd \tau$.  
As a result, we obtain 
the following practical scheme for integration of Eq.~(\ref{eq1}).
First, we pre-compute the coefficients $C_{1,2,3}$:
\[
C_1=\ii\eta^{-1}e^{-\gamma\Delta}(I_2\Delta-I_3)/2\Delta^2\;,\quad
C_2=\ii\eta^{-1}(I_3/\Delta^2-I_1)\;,\quad
C_3=-\ii\eta^{-1}e^{\gamma\Delta} (I_2\Delta+I_3)/2\Delta^2\;,
\]
where
\[
I_1=i\eta^{-1}(e^{-\ii\eta\Delta}-1)\;,\quad 
I_2=\eta^{-2}[e^{-\ii\eta\Delta}(1+\ii\Delta\eta)-1]\;, \quad 
I_3=\eta^{-3}[e^{-\ii\eta\Delta}(\Delta\eta(2+i\Delta\eta)-2\ii)+2\ii]\;.
\]
Next, for given $x_k,\dot x_k$ and the new measurement $s_{k+1}$ we compute   
$x_{k+1},\dot x_{k+1}$ in three steps: 
\begin{enumerate}
\item Compute $A_k=x_k-\ii(\dot x_k+\gamma x_k)/\eta$.
\item Compute $A_{k+1}=A_k+C_1s_{k-1}+C_2s_k+C_3s_{k+1}$.
\item Compute 
$
x_{k+1}=\mbox{Re}(A_{k+1}e^{\ii\eta\Delta})e^{-\gamma\Delta}\;, \qquad 
\dot x_{k+1}=\left [0.5\ii \eta (A_{k+1}e^{\ii \eta \Delta}-A^*_{k+1}e^{-\ii \eta \Delta})-\gamma\mbox{Re}(A_{k+1}e^{\ii\eta\Delta})\right] e^{-\gamma\Delta}\;.
$\\
We emphasize that all coefficients can be pre-computed, so that 
the integration step requires only a few summations and multiplications. 
\end{enumerate}

The extension of the approach to solving Eqs.~(\ref{eq:filter},\ref{eq:integrator}) 
of the resonant oscillator is straightforward. The equation~(\ref{eq:filter}) of the linear oscillator is solved as just described.
Then the known points $\dot x_{k-1},\dot x_k, \dot x_{k+1}$ provide the local parabolic approximation of the function $\dot x$ in Eq.~(\ref{eq:integrator}). 
The solution of this linear equation is readily
obtained by variation of the constant and reads:
\[
 z_{k+1}=(z_k-a+b\mu-2c\mu^2)e^{-\Delta/\mu}+
 a-b\mu +2c\mu^2+b\Delta-2c\mu\Delta+c\Delta^2 \;,
\]
with $a=\dot x_k$, $b=(\dot x_{k+1}-\dot x_{k-1})/2\Delta$, 
$c=(\dot x_{k-1}-2\dot x_k+\dot x_{k+1})/2\Delta^2$.

\bigskip
To summarize the presentation of numerical techniques, we emphasize that if
the sampling rate is very high, the differential equations in all presented algorithms can 
be fast and easily solved by the midpoint or predictor-corrector technique~\cite{Press_et_al-86}. 
However, the stability of these simple methods is not guaranteed. 

\subsection*{Choosing parameters}
\paragraph{Locking-based oscillator.} The main parameter is the forcing coefficient $\e$
 (Eq.~\ref{eq:phosc}). 
The larger $\e$ the faster the system synchronizes. On the other hand, the forcing term $\e\sin\theta\cdot s(t)$
shall be small enough to ensure the monotonicity of the phase $\theta$. So, if $s(t)=a\cos\vp$ then 
(Eq.~\ref{eq:phosc}) reads $\dot\theta=\w+\frac{\e a}{2}\sin(\vp-\theta)- \frac{\e a}{2} \sin(\vp+\theta)$;
in the locked state $\vp\approx\theta$ the first sine term vanishes and $\dot\theta>0$ if $\e a < 2\w$.
In the artificial data example we checked that he values $0.4\le \e\le 0.8$ provide good result. In this example, the amplitude goes up to $\approx 2$ and 
the frequency is about 1, so that the above condition is satisfied. The parameter of the frequency adaptation shall be  $K\le 1$; the values between 0.5 and 1 work well.  The data in
Fig.~\ref{fig:1} correspond to $\e=0.8$, $K=1$.
For the tremor data
$\w\approx2\pi\cdot4.5$,  $\e=30$, and $K=0.5$. In both cases we updated the frequency $\w$ 20 times 
per period using the preceding phases $\theta$ in the interval that is about one cycle long.

\paragraph{Non-resonant oscillator.}  To process the artificial signal we exploited 
$\alpha_a=6$ (amplitude measurement), $\alpha_p=0.2$ (phase measurement).
For the  LFP example, the parameters were  $\alpha_a=80$, $\alpha_p=10$. In both cases 
we took $\w=5\nu$, where $\nu=1$ and $\nu=2\pi\cdot 17$, respectively.
Practically, we have to choose $\alpha_p$ so that $|\tan\beta|\approx |\beta|\ll 1$.
For the common choice $\w=5\nu$ it reduces to the condition 
$|\beta|=\alpha_p/24\nu\ll 1$. (Notice that the smaller $\alpha$ the longer the transient.) Similarly, we have to choose $\alpha_a$ so that the derivative of the resonance curve 
$\left .\frac{d}{d\nu}(a/b)\right |_{\nu=\w/5}=
\left .\frac{2(\w^2-\nu^2)\nu-\alpha_a^2\nu}{[(\w^2-\nu^2)^2+\alpha_a^2\nu^2]^{3/2}}\right |_{\nu=\w/5}\approx 0$ what yields $\alpha_a\approx 7\nu$. For a bandpass filtered signal, the algorithm is not very sensitive to $\alpha_a$. So, for $\alpha_a\approx 0.75\nu$  in the LFP example, variation of the resonance curve within the bandpass is less than 4\%.

\paragraph{Resonant oscillator.} Here we always take $\alpha=0.3\w$ what corresponds to 
bandpass width about $30\%$ of the central frequency $\w$. 
The requirement for the parameter $\mu$ is $\mu\gg 1$; in all computations we choose 
$\mu=500$. 

\section*{Acknowledgements}
We thank Michael Feldman for useful discussions.
The work was supported by Deutsche Forschungsgemeinschaft 
(Project-ID 424778381 – TRR 295 Retune).

\section*{Author contributions statement}

M.R. and A.P. designed the algorithms. M.R. performed the computations, prepared the final figures, and wrote the manuscript with the contribution and corrections from all other authors. 
J.L.B. contributed LFP and accelerometer data.
All authors participated in the evaluation and discussion of the results.

\section*{Additional information}

The authors declare no competing interests. The Matlab scripts for all algorithms can be obtained from the authors 
upon request.


\end{document}